# Capabilities of the GAMMA-400 gamma-ray telescope to detect gamma-ray bursts from lateral directions

A A Leonov[1,2,*], A M Galper[1,2], N P Topchiev[1], I V Arkhangelskaja[2], A I Arkhangelskiy[2], A V Bakaldin[1,3], I V Chernysheva[2], O D Dalkarov[1], A E Egorov[1], M D Kheymits[2], M G Korotkov[2], A G Malinin[2], A G Mayorov[2], V V Mikhailov[2], A V Mikhailova[2], P Yu Minaev[1,4], N Yu Pappe[1], P Picozza[5,6], R Sparvoli[5,6], Yu I Stozhkov[1], S I Suchkov[1] and Yu T Yurkin[2]

[1]Lebedev Physical Institute, Moscow, 119991, Russia
[2]National Research Nuclear University «MEPhI», Moscow, 115409, Russia
[3]Scientific Research Institute for System Analysis, Moscow, 117218, Russia
[4]Space Research Institute, Moscow, 117997, Russia
[5]Istituto Nazionale di Fisica Nucleare, Sezione di Roma 2, Rome, Italy
[6]Physics Department of University of Rome Tor Vergata, Rome, Italy

*e-mail: leon@ibrae.ac.ru

**Abstract**. The currently developing space-based gamma-ray telescope GAMMA-400 will measure the gamma-ray and electron + positron fluxes using the main top-down aperture in the energy range from ~20 MeV to several TeV in a highly elliptic orbit (without shading the telescope by the Earth and outside the radiation belts) continuously for a long time. The instrument will provide fundamentally new data on discrete gamma-ray sources, gamma-ray bursts (GRBs), sources and propagation of Galactic cosmic rays and signatures of dark matter due to its unique angular and energy resolutions in the wide energy range. The gamma-ray telescope consists of the anticoincidence system (AC), the converter-tracker (C), the time-of-flight system (S1 and S2), the position-sensitive and electromagnetic calorimeters (CC1 and CC2), scintillation detectors (S3 and S4) located above and behind the CC2 calorimeter and lateral detectors (LD) located around the CC2 calorimeter.
In this paper, the capabilities of the GAMMA-400 gamma-ray telescope to measure fluxes of GRBs from lateral directions of CC2 are analyzed using Monte-Carlo simulations. The analysis is based on off-line second-level trigger construction using signals from S3, CC2, S4 and LD detectors. For checking the numerical algorithm the data from space-based GBM and LAT instruments of the Fermi experiment are used, namely, three long bursts: GRB 080916C, GRB 090902B, GRB 090926A and one short burst GRB 090510A. The obtained results allow us to conclude that from lateral directions the GAMMA-400 space-based gamma-ray telescope will reliably measure the spectra of bright GRBs in the energy range from ~10 to ~100 MeV with the on-axis effective area of about 0.13 m$^2$ for each of the four sides of CC2 and total field of view of about 6 sr.

# 1. Introduction

The GAMMA-400 gamma-ray telescope (Galper et al., 2018; Topchiev et al., 2019; Topchiev et al., 2016; Topchiev et al., 2016a) is being designed to achieve a broad range of scientific goals. These scientific goals are the following: precise searching for high-energy gamma-ray emission from annihilation or decay of dark matter particles; searching for new and studying known Galactic and extragalactic point and extended sources of high-energy gamma rays (supernova remnants, pulsars, binaries, microquasars, active galactic nuclei, blazars, quasars, etc.); studying isotropic and Galactic diffuse backgrounds; detection of gamma-ray bursts (GRBs) and other transients, as well as precise measuring electron + positron cosmic-ray fluxes (Leonov et al., 2019).

The key experimental goal of the GAMMA-400 gamma-ray telescope is to measure gamma-ray fluxes in the energy range from ~20 MeV to ~1 TeV with high angular and energy resolutions. The main top-down aperture of the instrument is about ±45° (Topchiev et al., 2019a).

The main preliminary planned targets for continuous long time (up to 100 days) source observation mode with GAMMA-400 are sources of the Galactic plane and bulge in contrast to the uniform scanning mode of Fermi-LAT, when the observation time of sources is only ~15% of the Fermi-LAT operation time (Abdollahi, et al., 2020). With this observation mode the main aperture covers solid angle ~$0.3\pi$. Taking into account that GRBs are isotropically detected using lateral aperture allows us to increase the field of view for GRB detection up to $2\pi$. In this paper, the additional capabilities of the GAMMA-400 gamma-ray telescope for GRB detection using the lateral aperture in the energy range from ~10 MeV to several hundred MeV are analyzed using Monte-Carlo simulations with GEANT4 software.

"Classical" high variable prompt emission of GRBs is seen mostly in the sub-MeV range as superposition of short FRED-like (Fast Rise and Exponential Decay) pulses. More than 13 000 bursts (http://www.ssl.berkeley.edu/ipn3/masterli.html) were detected and investigated in details in this range, giving the high level of statistics. In the energy range above 100 MeV, emission of GRBs is relatively well studied also (a couple hundreds of detections) by several experiments (e.g. Fermi/LAT and CGRO/EGRET). This high-energy component is presumably associated with the propagation of a relativistic jet, which is observed mainly at high energies above 100 MeV (Joshi & Razzaque, 2021). However, the nature of the high-energy emission and its possible connection with the low-energy classical gamma-ray emission is still debatable. There are several signs of at least bimodal behavior of high-energy emission (separate emission mechanism vs extension of low-energy emission to high energy).

The energy range of 10-100 MeV, being the transition zone between GRB components with energies less than ~MeV and more than 100 MeV, was weakly studied (Ajello et al., 2019). This is due to both the features of scintillation detectors used to observe GRBs (insufficient efficiency of photon detection in this range), and the shape of the energy spectrum of events in this range (usually described by power-law model with an index of $\alpha \sim -2$ and steeper). At the same time, this range is of particular interest, since, as already mentioned, some GRBs are characterized by a transition from the spectral component, corresponding to the pulses of the burst prompt emission (observed mainly in the sub-MeV range), to the additional high-energy extended duration component. In this view, investigation of GRBs and accumulation of their recording statistics by the GAMMA-400 lateral aperture could shed light on the problem of possible connection between high and low energy gamma-ray emission of GRBs and their physical nature.

The paper is organized as follows. The next section presents the schematic and physical description of the GAMMA-400 gamma-ray telescope. The simulation model is described in the third section. To detect reliably gamma rays from GRBs using the lateral aperture, the special selection algorithm has to be developed decreasing the background of charged particles and diffuse gamma rays. In the fourth section, the construction of such algorithm (off-line second-level trigger) is described, using the results of source and background simulations. Also in this section, the acceptance and on-axis effective area of the instrument for the lateral aperture, taking into account the off-line second-level trigger selection, are provided. Then it is necessary, applying the processed algorithm, to check the possibility of GRB spectra reconstruction. To test this point, the data from space-based GBM and LAT instruments of Fermi experiment are used. The description of experimental data for three long bursts (GRB 080916C, GRB 090902B, GRB 090926A) and one short burst (GRB 090510A) (Ajello et al., 2019) and the results of GRB spectra reconstruction are reported in the fifth section. Some remarks are placed in the conclusion.

## 2. The physical scheme of the GAMMA-400 gamma-ray telescope

The GAMMA-400 instrument consists of the anticoincidence system (AC top, AC lat), the converter-tracker (C), the time-of-flight (ToF) system from the detectors S1, S2, the position-sensitive calorimeter (CC1), electromagnetic calorimeter (CC2), the detectors S3 and S4 located above and behind the CC2 calorimeter and lateral detectors (LD) located around the CC2 calorimeter. GAMMA-400 is able to measure high-energy gamma rays on event-by-event basis using main top-down (from the top directions) and additional (from the lateral directions) apertures. The current physical scheme of the GAMMA-400 gamma-ray telescope is shown in Fig. 1. In the future, this scheme can be slightly modified depending on the type of carrier rocket and the final payload mass.

For the main top-down aperture, the dimensions and location of plastic scintillation detectors S1 and S2, which are separated by approximately 500 mm, determine the maximum gamma-ray telescope field of view (FoV), which is about ±45°. The converter-tracker consists of 13 layers of double (x, y) coordinate detectors made of scintillation fiber (SciFi) detectors. Each plane contains two groups of scintillating fibers, oriented in orthogonal directions with 4096 SiPM readout channels on each side. The size of pitch is 250 μm. The converter-tracker planes have an active area of ~ 1000 × 1000 mm$^2$. The first seven layers are interleaved with tungsten conversion foils with thickness of 0.1 $X_0$ ($X_0$ is the radiation length), next four layers are interleaved with tungsten conversion foils with thickness of 0.025 $X_0$ and final two layers have no tungsten. Using the four layers with thickness of 0.025 $X_0$ allows us to measure gamma rays down to 20 MeV. In this case, the on-line gamma-ray trigger for the energy range of 20–100 MeV and 100 MeV – 1000 GeV is the same: $\overline{AC} \times ToF$ (the absence of a signal in AC and the presence of ToF time signal from S1 and S2 for top-down directions only). The total converter-tracker thickness is ~1 $X_0$. The two-part calorimeter (CC1 and CC2) measures the particle energy. The CC1 imaging calorimeter consists of the plane from CsI(Tl) crystals and a layer of double (x, y) SciFi coordinate detector without tungsten the same as final two layers in converter-tracker. The CC2 electromagnetic calorimeter consists of 22×22 CsI(Tl) crystals. Each crystal has a dimension of 36×36×300 mm$^3$. The thickness of CC1 and CC2 is 2 $X_0$ and 16 $X_0$, respectively. The total calorimeter thickness for vertical incidence direction is 18 $X_0$ or ~0.9 $\lambda_0$ ($\lambda_0$ is the nuclear interaction length). Using the deep calorimeter allows us to extend the energy range up to several TeV for gamma rays and to reach an energy resolution ~2% at 100 GeV (Topchiev et al., 2019a). The total calorimeter thickness for lateral detection of particles is ~42 $X_0$ or ~1.95 $\lambda_0$, providing the possibility to measure intensities of high-energy electrons and gamma rays up to ~10 TeV.

The structure of SciFi converter-tracker with analog readout, ToF system with large flight distance (500 mm), and an additional layer of SciFi coordinate detector in CC1 allows us to obtain unprecedented angular resolution: ~0.01° at $E_\gamma$ = 100 GeV (Topchiev et al., 2019a).

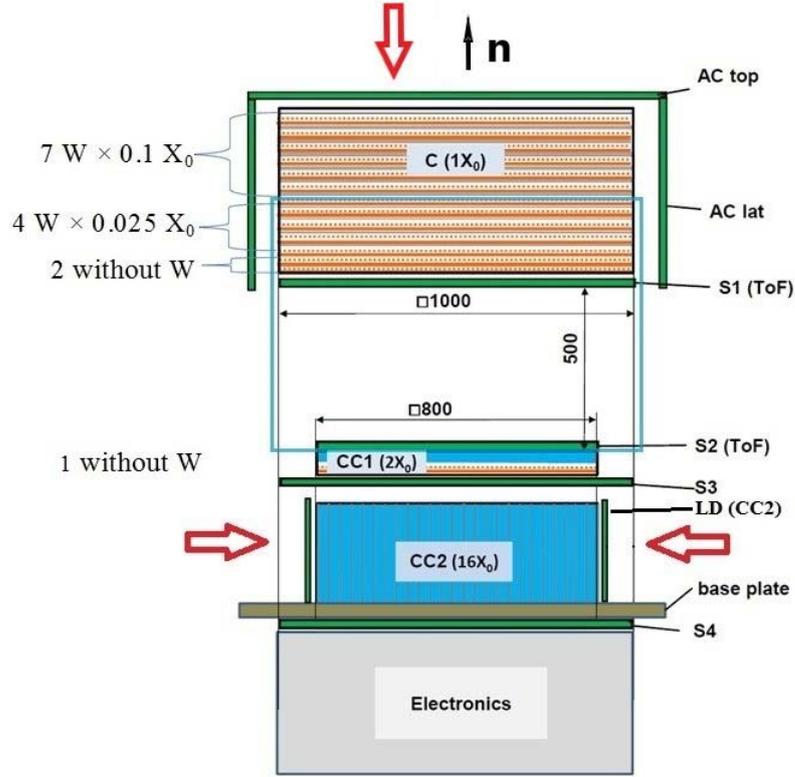

**Fig 1.** The current physical scheme of the GAMMA-400 gamma-ray telescope.

To detect gamma rays from lateral directions the information from CC2, S3, S4 and LD detectors is used. Each of the scintillation detectors (S3 and S4) consist of two layers of strips. Each strip has dimensions of 1000×100×10 mm$^3$ (length/width/thickness). LD detectors have the same structure with the strip dimensions of 380×117×10 mm$^3$.

The capabilities of the GAMMA-400 gamma-ray telescope for the lateral aperture are determined by the on-line special trigger signal to capture such gamma rays:

$$\overline{LD} \times \overline{S_3} \times \overline{S_4} \times CC_2(5\ MeV < E_{RELEASE} < 350\ MeV) \qquad (1),$$

where LD, S3, S4 are anticoincidence detectors; lower (5 MeV) and upper (350 MeV) limits for the signal in CC2 are conditioned by lower boundary of calorimeter dynamic range and the limitation of telemetric data volume, respectively.

One of the main problems for applying the GAMMA-400 lateral aperture to detect GRB gamma rays is to provide the effective rejection from all directions of background induced by cosmic-ray particles: protons, electrons, positrons, helium, and additional contribution induced by diffuse gamma rays. Charged particles are excluded by anticoincidence S3, S4, LD detectors and are able to be detected according criterion (1), if penetrate inside CC2 only through the holes between lateral detectors LD and S3, S4 scintillation detectors. At present, this mechanical structure with holes is due to the total mass limitation for the GAMMA-400 scientific payload and will be further modified.

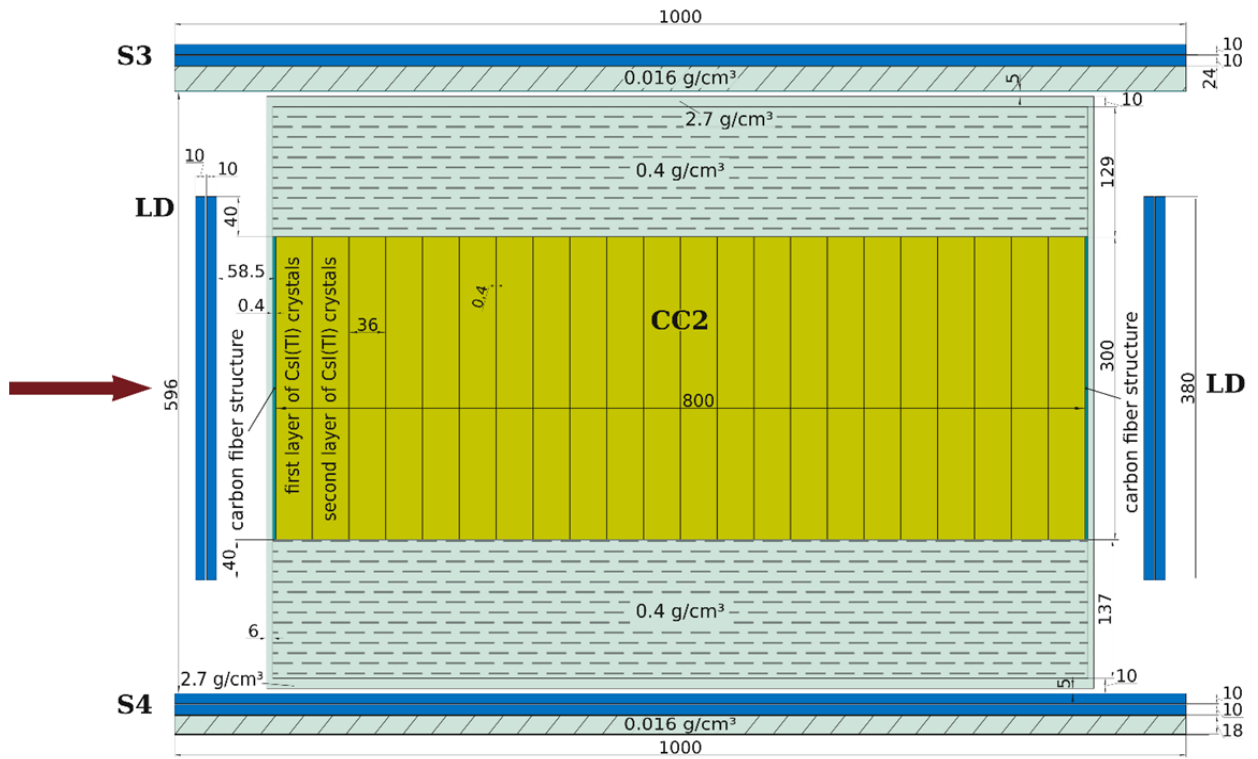

**Fig 2.** The part of the GAMMA-400 physical scheme, which is using to detect gamma rays from lateral directions.

The part of the GAMMA-400 physical scheme, which is using to detect gamma rays from lateral directions, is presented in Fig. 2 in details with the matter of supporting construction. Each of S3 and S4 detectors is installed on the honeycomb panel with average density 0.016 g/cm$^3$. The electromagnetic calorimeter CC2 is inserted in the supporting structure from aluminum and contains zones for installing electronics. Top part of this structure consists of the layer of aluminum (2.7 g/cm$^3$) and from the special module of stiffeners and electronics with effective density ~0.4 g/cm$^3$. The thickness of aluminum is 10 mm and the thickness of the special module is 129 mm. The bottom part of the supporting structure of CC2 contains the same special module of stiffeners and electronics with thickness of 137 mm and layer of aluminum with thickness 15 mm. Lateral parts from each side contain the layer of aluminum with thickness of 6 mm. CsI(Tl) crystal columns of CC2 are installed inside the cells of the carbon fiber honeycomb structure having thickness of 0.4 mm.

### 3. The simulation model

Using GEANT4 to evaluate the rejection power, simulations of the differential isotropic fluxes of cosmic-rays protons, electrons/positrons and helium and background of diffuse gamma rays were performed for whole coverage of the GAMMA-400 lateral aperture. The surfaces, over which the simulations were carried out, are shown in Fig. 3 by blue dashed lines.

When simulating, only five (top-down and four lateral) possible directions are considered. From bottom direction, the penetration of particles with logic trigger (1) is not considered due to large amount of satellite platform matter located under the gamma-ray telescope. A lot of shower particles, penetrating from bottom, will be excluded by the veto signal in S4.

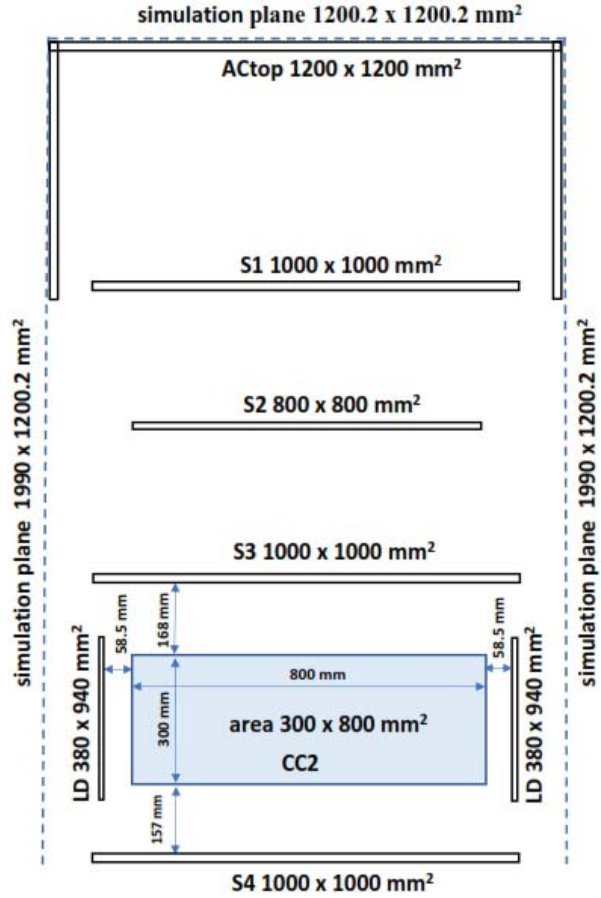

**Fig. 3.** The scheme of simulations.

To simulate cosmic-ray background, it is necessary to mention that due to the 6-mm thickness of the aluminum box from lateral sides, which surrounds the CC2 calorimeter, only particles with energies above the certain threshold can be captured according condition (1). This threshold is equal to 35 MeV for protons, 140 MeV for helium nuclei and 3.4 MeV for electrons/positrons. Taking into account this point for calculations, we use the experimental data of of the PAMELA experiment during solar minimum for protons (Martucci et al., 2018), electrons/positrons (Adriani et al., 2015) and helium (Adriani et al., 2011) were used. The low energy limits for charged particle detection in this experiment is about 80 MeV (for protons), 50 MeV (for electrons), 120 MeV/n (for helium) (Picozza et al., 2007). In low-energy range, the constant extrapolation for proton spectra is used. Such approach gives some overestimate for the influence of cosmic-ray background (Bisschoff et al., 2019). For electrons and helium in low-energy range, the estimations from (Bisschoff et al., 2019; Ngobeni et al., 2020) were used, respectively. These estimations are accounted for solar modulation effects. The cosmic-ray spectra used in simulations are shown in Fig. 4 by red (protons), green (helium) and blue (electrons) lines.

As a simple estimate of the diffuse gamma-ray background, we took the average intensity of the diffuse Galactic emission inside the disc with 30° radius around the Galactic center according to Fermi-LAT data (Ackermann et al., 2012; https://fermi.gsfc.nasa.gov/ssc/data/access/lat/BackgroundModels.html) and extrapolated such intensity isotropically to the whole sky. Fermi-LAT presents the data down only to 50 MeV, but we extrapolated the intensity linearly below this energy. We use such simple approach with evident overestimation, because the main contribution in the background comes from protons. The accuracy of this estimation does not influence on the result. We also checked that the contributions from point sources and isotropic background are much smaller than the Galactic component estimated this way. Hence, we neglected by these contributions. In general, the simulation results do not depend strongly from the chosen gamma-ray background model. The spectrum of diffuse gamma rays, used in simulations, is shown in Fig. 4 by black line.

For each spectrum ~$10^7$ events were simulated over the simulation planes, which are shown in Fig. 3.

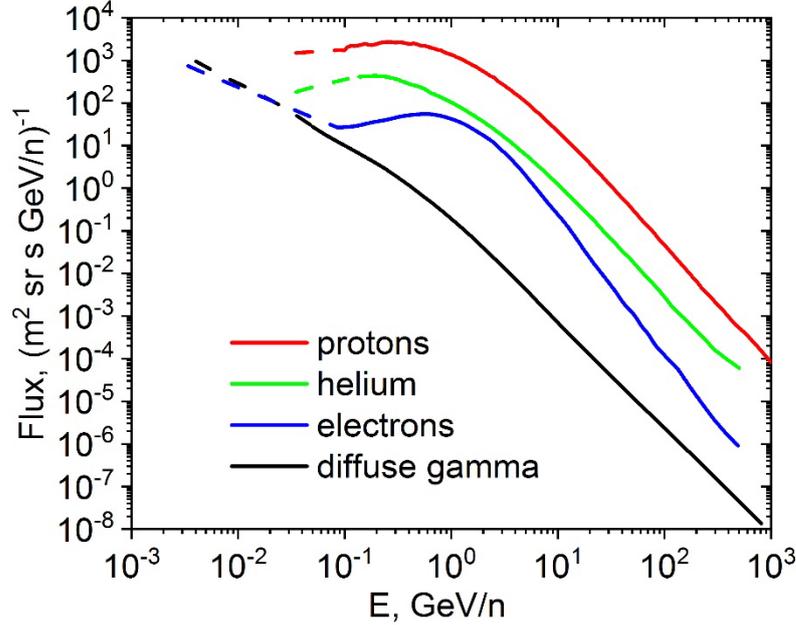

**Fig. 4.** The cosmic-ray spectra for protons (red) (Martucci et al., 2018), helium (green) (Adriani et al., 2011; Picozza et al., 2007), electrons (blue) (Adriani et al., 2015; Picozza et al., 2007) and of diffuse gamma-rays (black) (Ackermann et al., 2012; https://fermi.gsfc.nasa.gov/ssc/data/access/lat/BackgroundModels.html) used in the background simulations.

## 4. Off-line second-level trigger for background rejection.

To take into account a realistic description of the full measurement processes involved in a space-based scintillation detector, the results of calibration measurements (Suchkov et al, 2021) were applied into the simulations using approach, described in (Leonov et al., 2015).

Analyzing the results of simulation for the differential isotropic fluxes of cosmic-ray protons, electrons, positrons and helium and for differential flat flux of GRB emission the following estimations for criterion (1) were obtained. Protons provide count rate ~185 Hz; electrons + positrons and helium contribute ~5 and ~9 Hz, respectively; the contribution of emissions of considered GRBs begins from ~21 Hz. It is seen that applying on-board trigger selection (1) for the cosmic-ray particles and gamma-ray emission, which arrive at CC2 from all five directions, the dominant volume of the data, satisfied this selection, will be determined by cosmic-ray particles, in the main, by protons. To extract useful signal from captured gamma rays, additional criterions are needed. Using signals from CsI(Tl) crystals of CC2, it is possible to construct off-line second-level trigger logic to decrease significantly the count rate of cosmic-ray particles (Mikhailova et al., 2020). This logic is based on differences in electromagnetic and hadronic cascades in the CC2 calorimeter.

Considering for definiteness, gamma-ray emission, arriving from left direction of CC2, the significant probability of the conversion arises in first layer of CsI(Tl) crystals (Fig. 2). In this case, the energy is released in the nearest crystals. Then the electromagnetic shower is quickly weakened in calorimeter depth. Low-energy protons survived after applying criterion (1) and arriving to CC2 through the holes between lateral detectors LD and scintillation detectors S3 and S4, with high probability do not give a signals in first layer of CsI(Tl) crystal due to small solid angle of the holes. An energy release distribution on the layers of CsI(Tl) crystals for protons penetrating deeply inside the CC2 will be different in comparison the distribution for gamma rays with the same energy. Besides, protons are able to interact in second or third layers of CsI(Tl) crystals and to provide the location of energy release maximum deeply inside CC2. To take into account these differences gamma-ray and proton responses inside CC2, the following parameters are used:

$E_{CC_2}^{tot}$ is total energy release in all CsI(Tl) crystals of CC2;

$E_{CC_2}^1$ is total energy release in first layer of CsI(Tl) crystals of CC2;

$E_{CC_2}^2$ is total energy release in second layer of CsI(Tl) crystals of CC2;

$N_{hits}$ is number of CsI(Tl) crystals with energy release over threshold of 5 MeV;

$E_{CC_2}^{MAX}$ is maximum energy release in one of CsI(Tl) crystals;

$N_{layer}(E_{CC_2}^{MAX})$ is number of layer, which contains the crystal of CsI(Tl) with maximum energy release in CC2.

As example, the distribution of gamma-ray and proton events in $(E_{CC_2}^2, E_{CC_2}^{tot})$ phase space, satisfying condition (1), is presented in Figs. 5a and 5b, respectively. Criterion (1) keeps $3.8 \times 10^4$ protons and $4.6 \times 10^5$ gamma rays from $6.3 \times 10^6$ and $5.5 \times 10^6$ events, respectively, if isotropic fluxes from Fig. 4 are simulated on the left simulation plane (Fig. 3). Separation of gamma rays and protons is reached by introducing the linear cut (red line) corresponding to the expression $E_{CC_2}^2 = (0.5 \times E_{CC_2}^{tot} - 50)$. It provides the rejection of protons of 85.5% and the efficiency of gamma-ray detection of 51%.

As another example, the distribution of $N_{hits}$ parameter for gamma-ray (green columns) and proton (red columns) events, satisfying condition (1), with energy release $75\ MeV < E_{CC_2}^{tot} < 120\ MeV$ is shown in Fig. 6. Introducing a criterion, which is shown by blue dashed line, we obtain the rejection of protons of 74% and efficiency of gamma-ray detection of 89%.

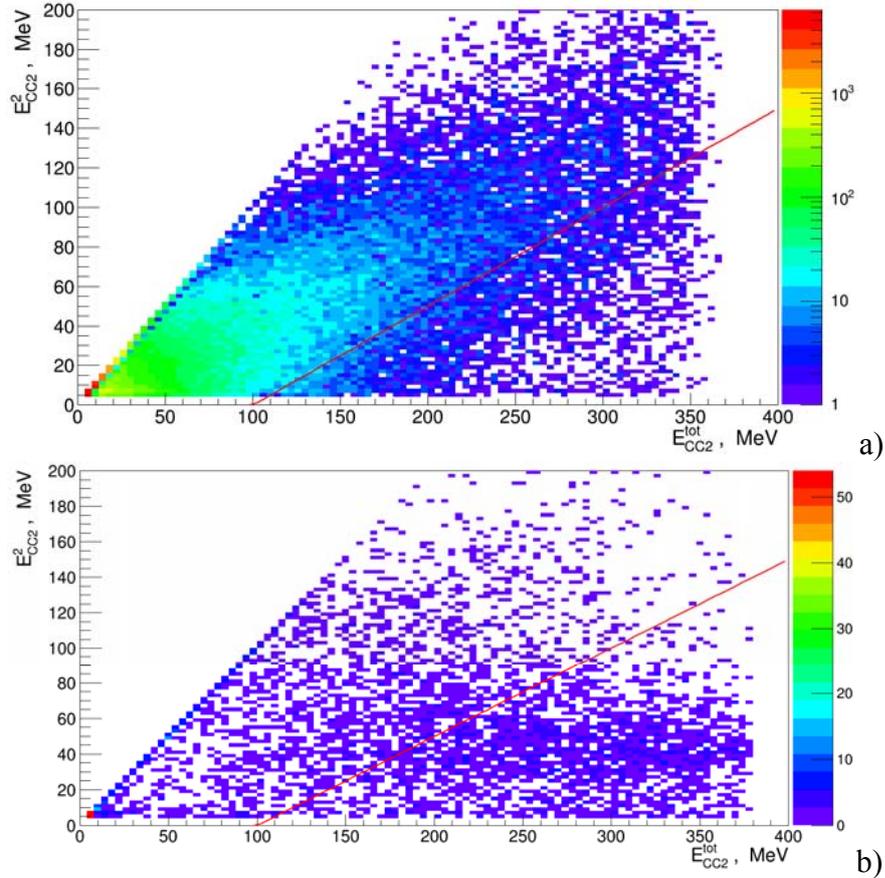

**Fig. 5.** The distribution of gamma-ray (a) and proton (b) events, satisfying condition (1), in $(E_{CC_2}^2, E_{CC_2}^{tot})$ phase space.

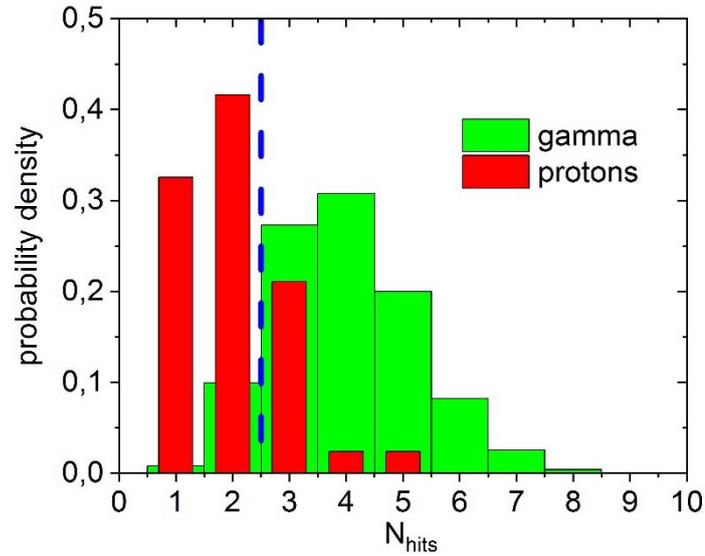

**Fig. 6.** The distribution of $N_{hits}$ parameter for gamma-rays (green columns) and protons (red columns) with energy release of $75\ MeV < E_{CC_2}^{tot} < 120\ MeV$. Blue dashed line denotes a criterion, which provides the rejection of protons with level 74% and efficiency of gamma-rays with level 89%.

To confirm the possibility to reject protons from gamma rays in the lateral aperture an independent analysis based on machine learning analyses was applied. Usually the particles identification, based on Multivariate analysis (MVA), gives the best performance, because such analysis allows to combine several discriminating variables into one final discriminator, takes into account the correlation between parameters and events, failing a particular criterion, are not ruled out if other criteria could classify them properly. Multivariate analysis of data is easy to understand and interpret with Decision Tree algorithm when consecutive set of questions with only two possible answers are applied to classify events. Final classification corresponds to leaf of the tree with appropriate weighting parameter.

The boosted decision tree (BDT) classifier was used from the toolkit TMVA included in CERN's ROOT package (Brun & Rademakers, 1997). Four parameters: $N_{hits}$, $E_{CC_2}^2$, $E_{CC_2}^{MAX}$ and $E_{CC_2}^1$ with a more significant discrimination power between gamma rays and protons together with released energy $E_{CC_2}^{tot}$ were combined into a single discrimination function. Using data of Monte-Carlo simulation both for gamma rays and protons, the BDT algorithm was trained with 10% of data for signals and background. Examples of BDT response distributions are shown in Fig. 7 for samples of gamma-ray and proton data. The data passed the trigger criterion (1). The BDT method provides better selection efficiency, but to achieve necessary energy resolution it has to be accompanied by additional cuts. Nonetheless, even in this simplified case, the location of distributions for gamma rays and protons in Fig. 7 confirms the possibility of rejection. For example, if the value of BDT parameter equal 0, then it provides gamma-ray efficiency of 90% and protons rejection of 87%.

Taking into account the results of the cosmic-ray and diffuse gamma-ray simulations (Fig. 4), the following additional criterions, providing effective rejection of charged particle background and suitable efficiency of gamma rays (~85 % in ~40 - 100 MeV energy interval), were selected:

$E_{CC_2}^{tot} > 7$ is introduced to increase the threshold for excluding indistinguishable events close to low-energy threshold;

$E_{CC_2}^1 > 99.21 - 249.5 \times e^{-\frac{E_{CC_2}^{tot}}{88.4}}$ is introduced to take into account that correlation between total energy release in first layer of CsI(Tl) crystals and total energy release in all CsI(Tl) crystals differs for proton and gamma-ray induced cascades (it begins from ~50 MeV);

$E^1_{CC_2} > 5$ corresponds to lower boundary of calorimeter dynamic range;

$$\begin{bmatrix} \left((N_{hits} = 1) \times (E^{tot}_{CC_2} < 52)\right) \big| \left((N_{hits} = 2) \times (E^{tot}_{CC_2} < 75)\right) \big| \\ \left((N_{hits} = 3) \times (E^{tot}_{CC_2} < 120)\right) \big| \left((N_{hits} = 4) \times (E^{tot}_{CC_2} < 150)\right) \big| (N_{hits} > 4) \end{bmatrix}$$ presents the difference of energy releases in CsI(Tl) crystals in proton and gamma-ray induced cascades;

$\left(N_{layer}(E^{MAX}_{CC_2}) \leq 4\right)$ uses the fact that in the considered energy range the position of shower maximum is limited by four layers, if it is initiated by gamma ray;

$E^2_{CC_2} > \left(0.5 \times E^{tot}_{CC_2} - 50\right)$ is introduced to take into account that correlation between total energy release in second layer of CsI(Tl) crystals and total energy release in all CsI(Tl) crystals differs for proton and gamma-ray induced cascades;

$E^{MAX}_{CC_2} < 250$ presents the difference of maximum energy release in CsI(Tl) crystals in proton and gamma-ray induced cascades.

The total combination of the developed criterions looks like:

$$(E^{tot}_{CC_2} > 7) \times \left(E^1_{CC_2} > 99.21 - 249.5 \times e^{-\frac{E^{tot}_{CC_2}}{88.4}}\right) \times (E^1_{CC_2} > 5) \times (E^1_{CC_2} < 200) \times$$
$$\begin{bmatrix} \left((N_{hits} = 1) \times (E^{tot}_{CC_2} < 52)\right) \big| \left((N_{hits} = 2) \times (E^{tot}_{CC_2} < 75)\right) \big| \\ \left((N_{hits} = 3) \times (E^{tot}_{CC_2} < 120)\right) \big| \left((N_{hits} = 4) \times (E^{tot}_{CC_2} < 150)\right) \big| (N_{hits} > 4) \end{bmatrix} \times \quad (2),$$
$$\left(N_{layer}(E^{MAX}_{CC_2}) \leq 4\right) \times \left(E^2_{CC_2} > \left(0.5 \times E^{tot}_{CC_2} - 50\right)\right) \times (E^{MAX}_{CC_2} < 250)$$

where energy releases are expressed in MeV.

The results of applying criterion (2) to the responses of CC2, when detecting spectra of cosmic rays and diffuse gamma rays according to Fig. 4, are presented in Table 1. In this table the rejection power of this criterion for the candidate events, satisfying conditions (1), is pointed out. It is seen that protons are rejected more than one order of magnitude stronger than gamma rays.

*Table 1. The rejection power of the criterion (2).*

| Particle type | Number of events (criterion (1) + criterion (2))/ Number of events (criterion (1)), % |
|---|---|
| Protons | 3.1 |
| Electrons/positrons | 6.6 |
| Helium | 2.1 |
| Gamma rays | 37.4 |

The GAMMA-400 acceptance and on-axis effective area for gamma-ray detection arriving from one lateral direction after applying of criterions (1) and (2) are shown in Figs. 8a and 8b, respectively. The values of on-axis effective area and acceptance reach ~0.13 m$^2$ for each lateral side and ~0.42 m$^2 \times$sr, respectively. The dependence of average total energy release in calorimeter CC2 from initial energy of gamma rays, arriving from one lateral direction and satisfying criterions (1) and (2), is presented in Fig. 9a. From this plot it is seen that appropriate linearity is provided. An example of energy release distribution for 50-MeV gamma rays, satisfying criterions (1) and (2), is shown in Fig. 9b.

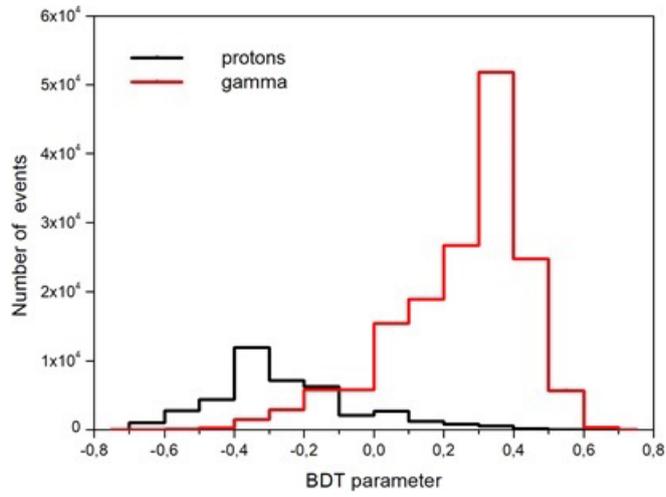

**Fig. 7**. Example of BDT classifier response distributions for samples of gamma-ray and proton data passed the trigger criterion (1).

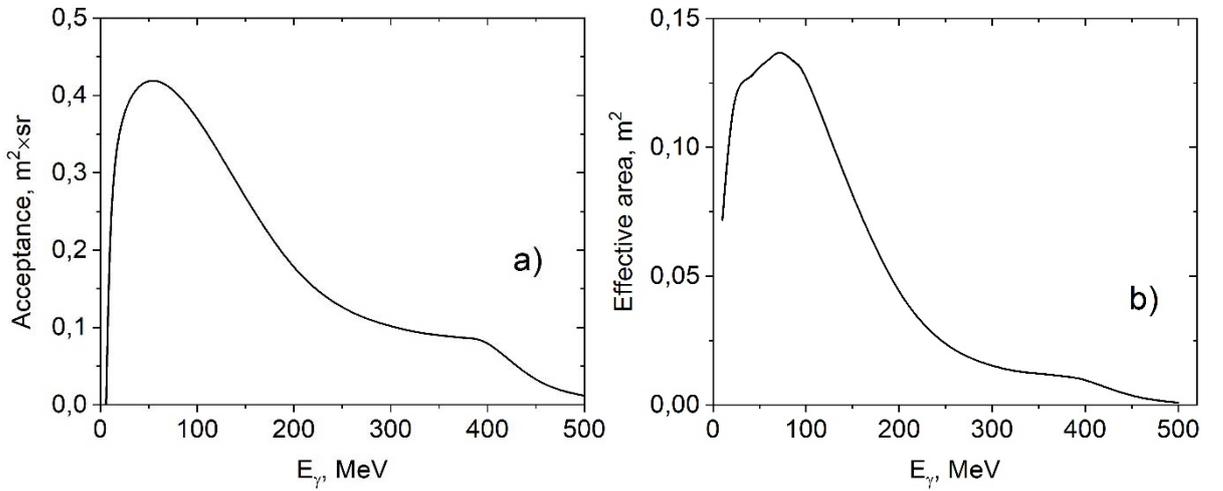

**Fig. 8.** The GAMMA-400 acceptance (a) and on-axis effective area (b) for gamma-ray detection arriving from one lateral direction after applying criteria (1) and (2).

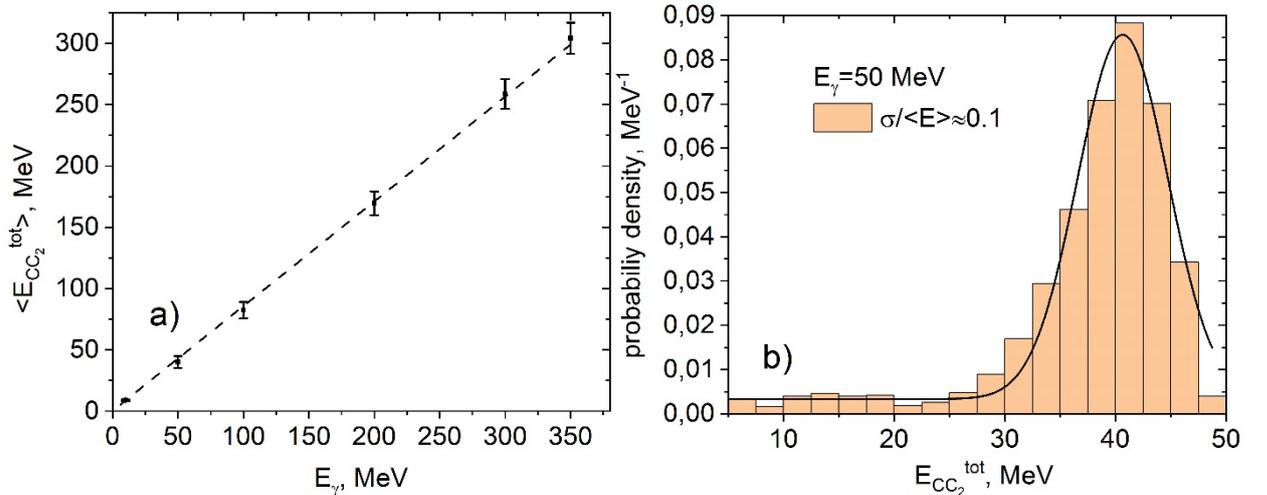

**Fig. 9.** The dependence of average total energy release in calorimeter CC2 from initial energy of gamma rays, arriving from one lateral direction and satisfying criteria (1) and (2) (a). Energy release distribution for 50-MeV gamma rays, satisfying criteria (1) and (2) (b).

## 5. The results of GRB spectra reconstruction

To restore the initial flux when detecting GRB emission, which arrives from left direction with normal incidence for definiteness, we use the GAMMA-400 performance from Figs. 8b and 9. The initial flux in energy bin $[E_i, E_i + \Delta E_i]$ can be obtained applying unfolding procedure according to the expression (An et al., 2019):

$$F(E_i, E_i + \Delta E_i) = \frac{N_{inc,i}}{S_{eff,i} \times \Delta T \times \Delta E_i}$$

$$N_{dep,j} = \sum_i M_{ji} N_{inc,i} \qquad (3),$$

where $N_{inc,i}$ is the number of events in the $i$th incident energy bin;
$N_{dep,j}$ is the number of events in the $j$th deposited energy bin;
$M_{ji}$ is the response matrix, i.e., the probability that an event in the $i$th incident energy bin is detected in the $j$th deposited energy bin;
$S_{eff,i}$ is the effective area;
$\Delta E_i$ is the width of the energy bin;
$\Delta T$ is the exposure time.

However, to make a conclusion about the possibility of GAMMA-400 instrument to detect reliably GRB emission through lateral directions, it is necessary to evaluate the background from cosmic rays and diffuse gamma rays. Applying the same algorithm (3), the contribution of background fluxes was calculated. To introduce the real time in the analysis of background simulation results the following expression is used:

$$S \int_{2\pi} \int_{E1}^{E2} I(E) \cos(\theta) d\Omega dE = \pi S \int_{E1}^{E2} I(E) dE = \frac{N_0}{\Delta t} \qquad (4),$$

where $I(E)$ is the background isotropic intensity of cosmic rays or diffuse gamma rays (Fig. 4);
$S$ is the surface area, over which the initial flux was simulated (left simulation plane in Fig. 3);
$E_1, E_2$ are the borders of energy range for given type of particles;
$N_0$ is number of simulated events;
$\Delta t$ is real time interval.

After applying of criterions (1) and (2) for $N_0$ simulated events, only $N_{BG}$ background events are remained. The effective, i.e., virtually flat with normal incidence, background flux in some energy bin $\Delta E_i$ is calculated from (3).

The results of calculations of background fluxes, which satisfy criterions (1) and (2), are presented in Fig. 10. In Fig. 10a, background fluxes from separate components are shown. In this figure, the contributions from top and four lateral directions are summed for each component. As already mentioned, the diffuse gamma-ray flux was estimated for direction to Galactic center and can be considered as upper limit. In Fig. 10b, there are background flux from charged particles and total background flux. The dominant contribution of charged particles in background flux is seen.

To check the possibility to measure GRB spectra from lateral directions, the simulation of differential plane flux of gamma-ray emission was fulfilled for the one (for example, left) simulation plane (Fig. 3). For this, we use the data from the GBM and LAT instruments of Fermi experiment: four brightest (in terms of the signal-to-noise ratio) gamma-ray bursts were selected (Ajello et al., 2019). Only for the brightest bursts it is possible to perform the full-band analysis so that the count statistics in the 5-100 MeV range is sufficient (the number of counts from the source significantly exceeds the number of counts from the background signal). Among the investigated GRBs, the sample contains three long bursts (GRB 080916C, GRB 090902B, GRB 090926A) and one short one (GRB 090510A). The analysis used data from the most illuminated detectors (selected individually for each burst) of two types: based on NaI (energy range of 7-900 keV) and BGO (energy range of 200 keV - 45 MeV) crystals from the GBM experiment, as well as data from the LAT experiment in the low-energy mode (energy range of 20 MeV - 1 GeV). The procedure of the construction and fitting for the energy spectra of selected GRBs is described in section 5.

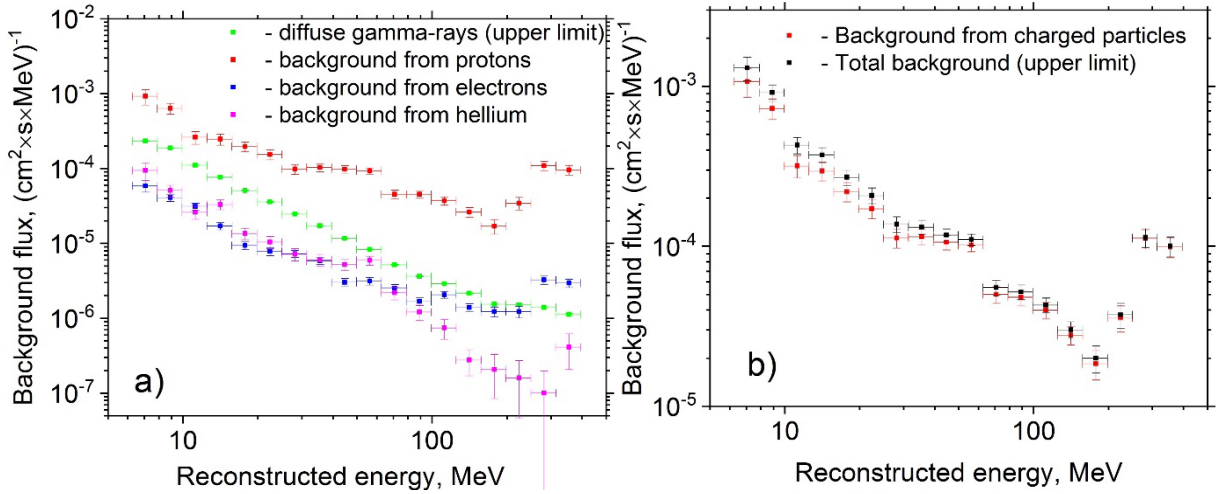

**Fig. 10.** The results of calculations of the effective background fluxes, which satisfy criteria (1) and (2). Background fluxes from separate components (a); background flux from charged particles and total background flux (b).

To construct and fit the energy spectra, the RMfit v4.3.2 software package, specially developed for the analysis of GBM and LAT/LLE data from the Fermi observatory (http://fermi.gsfc.nasa.gov/ssc/data/analysis/rmfit/), was used. The spectral analysis method is similar to that proposed in (Gruber et al., 2014), where the RMfit software package was also used. The energy spectra are approximated in the RMfit software package with the highest possible energy resolution, therefore the number of counts in each energy channel is small and, as a consequence, cannot be described by a normal distribution. Therefore, to approximate the energy spectra and to select the optimal spectral model, instead of the $\chi^2$ criterion, modified Cash statistics was used (CSTAT) (Cash, 1979). We used following spectral models (Poolakkil et al., 2021) and their combinations to fit energy spectra and to choose optimal model for each spectrum individually: power law (PL), power law with exponential cutoff (CPL), the Band GRB function, thermal model (BB). The constructing of energy spectra was performed in time intervals, covering the light curve peaks (two peaks for GRB 090926A) and the whole burst. The results of spectral analysis (optimal spectral model for each GRB) are presented in table 2. It was found that the energy spectrum of GRB 080916C (both integrated and in the peak region of the light curve) is described by Band GRB function in the entire spectrum range (7 keV - 1 GeV), indicating that high-energy emission has the same nature as low energy. A similar behavior was found for the first episode of GRB 090926A. In the remaining cases considered, high-energy emission represents a separate component, which is described by either a simple power-law model (PL) or Band GRB function.

Analyzing the spectrum for the whole burst gave a possibility to verify the sensitivity of GAMMA-400 lateral aperture to detect this burst, and analyzing the peak spectrum allowed to verify the possibility of looking for the details of spectrum using the GAMMA-400 performance.

We constructed the light curves in energy ranges from 20 MeV to 1 GeV and from 7 keV to 20 MeV for each of considered GRBs using initial data for the GBM and LAT experiments of Fermi available from the public FTP archive (ftp://legacy.gsfc.nasa.gov/fermi/data/). The light curves are shown in Fig. 11. Vertical lines denote time intervals, over which the experimental data were averaged, while constructing spectrum. For GRBs: 080916C, 090902B and 090510A two time ranges were considered, while for GRB 090926A three time ranges were used.

*Table 2. The results of spectral analysis of GRBs based on Fermi/LAT and Fermi/GBM data.*

| GRB | Time interval, s | Spectral model | Amplitude, $\frac{ph}{cm^2 \times s \times keV}$ | alpha | beta | Ep, keV |
|---|---|---|---|---|---|---|
| 080916C | (-1, 70) | Band | 0.0167+/-0.0003 | -1.04+/-0.01 | -2.18+/-0.01 | 498+/-20 |
|  | (4, 7) | Band | 0.0296+/-0.0006 | -1.17+/-0.02 | -2.23+/-0.04 | 2456+/-360 |
| 090510A | (0, 1) | Band+PL | 0.018/-0.002 | -0.66+/-0.06 | -3.3+/-0.2 | 3990+/-290 |
|  |  |  | 0.0039+/-0.0012 | -1.56+/-0.04 |  |  |
|  | (1, 9) | Band | 0.0010+/-0.0002 | -1.44+/-0.04 | -2.2+/-0.2 | (8.5+/-3.0)×10$^4$ |
| 090902B | (0, 25) | Band+PL | 0.0928+/-0.0006 | -0.59+/-0.01 | -3.51+/-0.11 | 752 +- 10 |
|  |  |  | 0.0129+/-0.0003 | -1.92+/-0.01 |  |  |
|  | (8, 10) | CPL+PL | 0.073+/-0.002 | 0.12+/-0.04 |  | 947+/-21 |
|  |  |  | 0.052+/-0.001 | -1.95+/-0.01 |  |  |
| 090926A | (-1, 21) | Band+CPL | 0.043+/-0.016 | -1.2+/-0.1 | -2.15+/-0.02 | 640+/-270 |
|  |  |  | 0.102+/-0.017 | -0.37+/-0.09 |  | 288+/-7 |
|  | (3, 6) | Band | 0.302+/-0.005 | -0.58+/-0.01 | -2.43+/-0.02 | 373+/-7 |
|  | (9, 11) | Band+CPL | 0.032+/-0.003 | -1.71+/-0.03 | -2.07 +- 0.11 | (2.9+/-2.2)×10$^4$ |
|  |  |  | 0.278+/-0.008 | -0.44+/-0.06 |  | 263+/-6 |

The initial spectrum from GRB 080916C, covered the light peak, averaged over 3-s time interval and used in simulations, is shown in Fig. 12 by red line. The estimated flux from GRB 080916C obtained using (3) is presented in Fig. 12a by red squares. In each logarithmic decade, ten equal energy bins were used. In each energy bin, the number of simulated events was corrected according to real time of 3 s (Fig. 12a) using the following expression:

$$\int_{E1}^{E2} I_{GRB}(E) \times S_{eff}(E) \times dE = \frac{N_0}{\Delta t} \quad (5),$$

where $I_{GRB}(E)$ is flat flux from GRB.

Besides, in Fig. 12a there are total background and observed (background + GRB) fluxes denoted by black squares and blue circles, respectively. The total background flux was constructed from simulation events applying time normalization (4) for 3-s time interval. Finally, the reconstructed flux from GRB was deduced as subtraction of background from the observed flux. The idea is the following. The total background flux is well measured and the observed flux is varied due to the additional contribution from GRB only. The reconstructed GRB flux is presented in Fig. 12b by red squares. From this figure it is seen that GRB 080916C source can be reliably measured with 3-s time bin in the energy range from ~10 MeV to 100 MeV using lateral aperture. To avoid uncertainties, concerning sharp profile of effective area (Fig. 8b) and the influence of measurement errors on empirical parameters in criterion (2), the following conclusions are considered for the initial energy of GRB emission > 10 MeV.

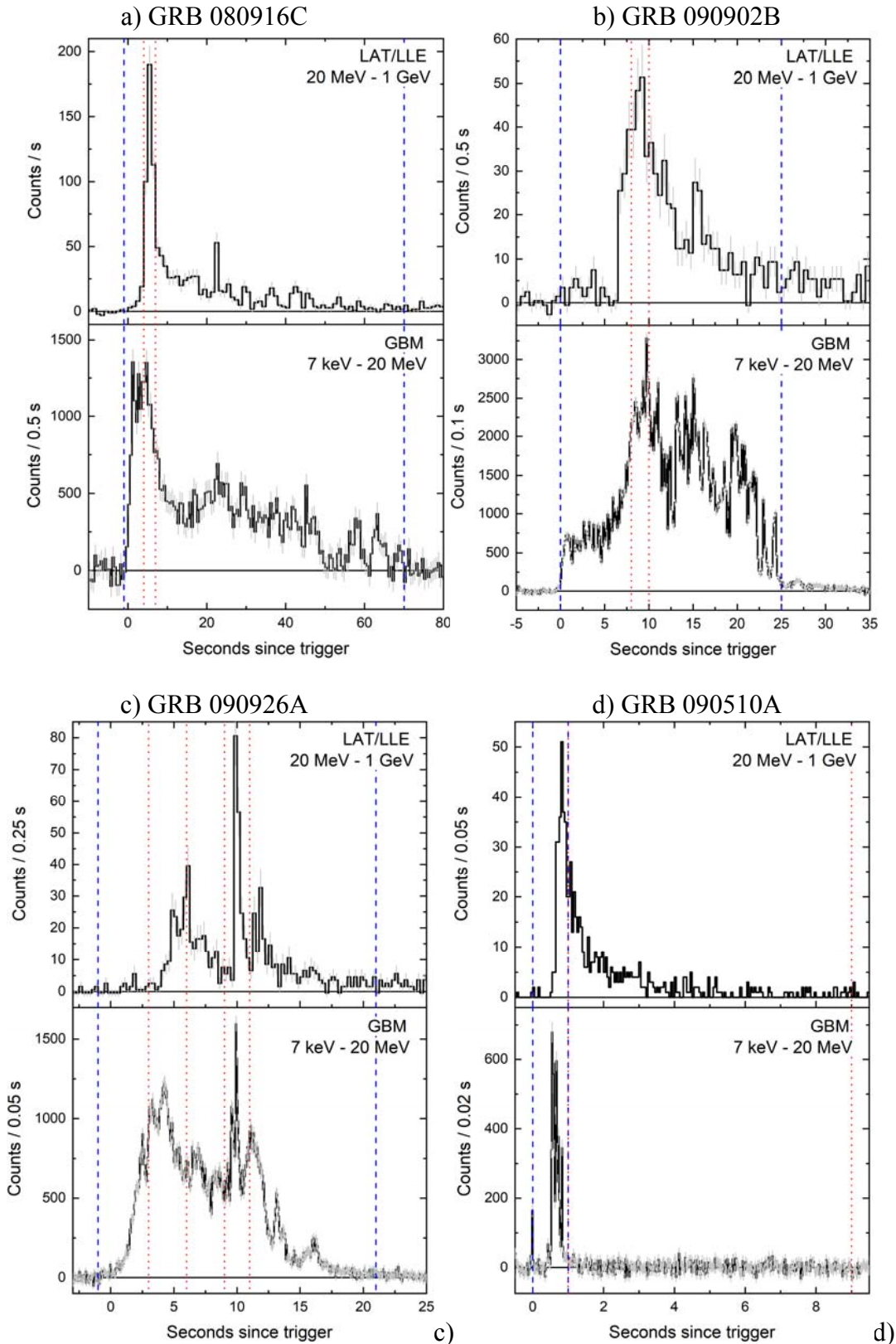

**Fig. 11.** Light curves for GRB 080916C (a), GRB 090902B (b), GRB 090926A (c), GRB 090510A (d) constructed using initial data for the GBM and LAT experiments of Fermi available from the public FTP archive (ftp://legacy.gsfc.nasa.gov/fermi/data/). Vertical lines denote time intervals, over which the experimental data were averaged.

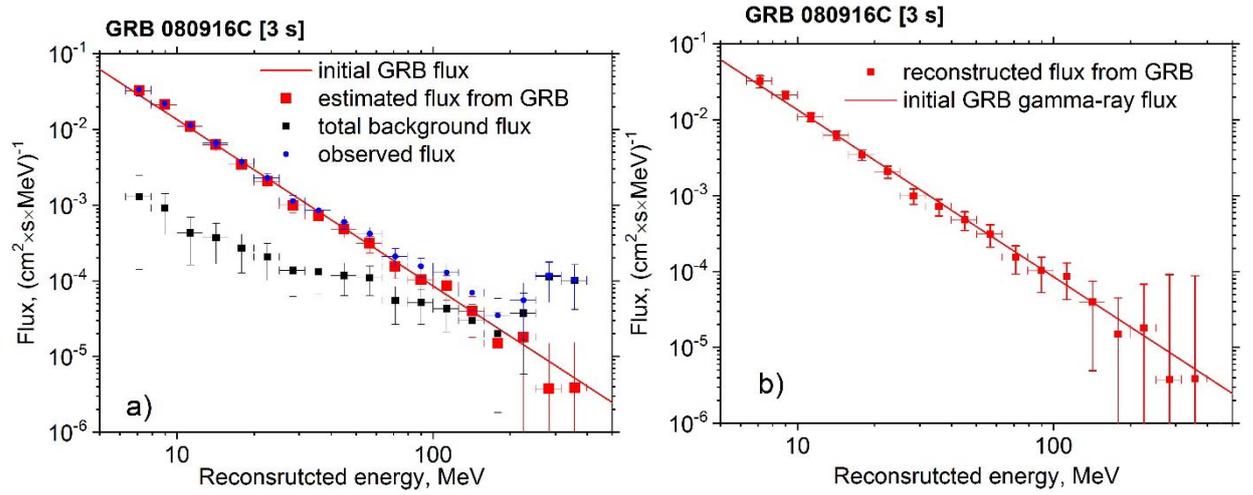

**Fig. 12.** The initial spectrum from GRB 080916C, averaged over 3-s time interval (red line); the estimated flux from GRB 080916C (red squares); the total background flux (black squares); the observed flux (background + GRB, blue circles) (a). The reconstructed flux from GRB 080916C (red squares) (b).

The initial spectrum from GRB 080916C, covered the whole burst, averaged over 71-s time interval and used in simulations, is shown in Fig. 13 by red line. The estimated flux from GRB 080916C, obtained by numerical algorithm, is presented in Fig. 13a by red squares. In each energy bin, the number of simulated events was corrected according to real time of 71 s (Fig. 13a).

Also, in Fig. 13a there are total background and observed (background + GRB) fluxes denoted by black squares and blue circles, respectively. The total background flux was constructed from simulation events applying time correction (3) for 71-s time interval. Finally, the reconstructed flux from GRB was deduced as subtraction of background from the observed flux. The reconstructed GRB 080916C flux, covered the whole burst, is presented in Fig. 13b by red squares.

From the Figs. 12b and 13b it is seen that GRB 080916C can be reliably measured as with time interval covered the peak, as with time interval covering the whole burst in the energy range from ~10 MeV to 100 MeV using lateral aperture detection.

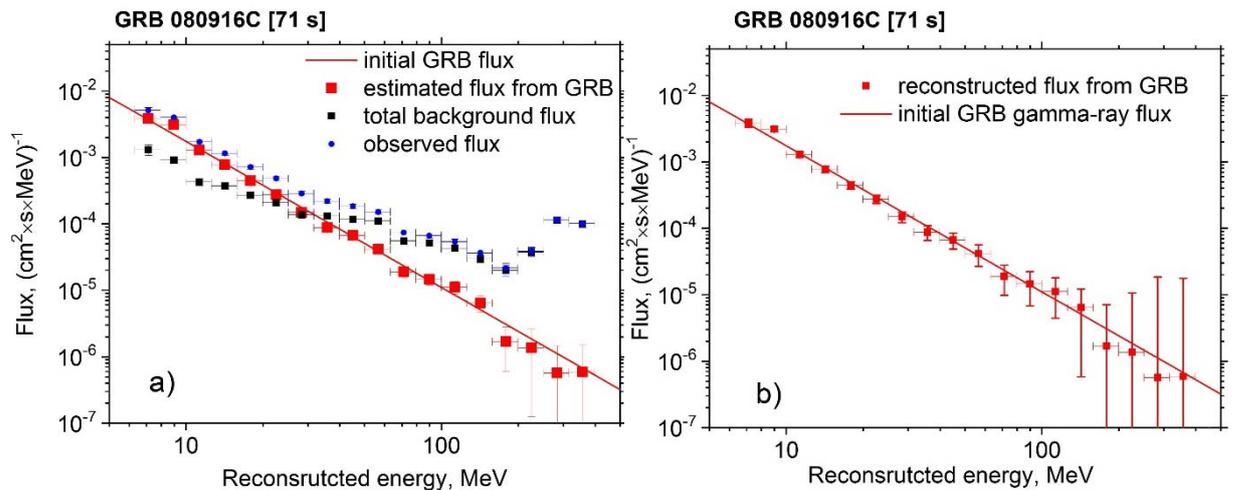

**Fig. 13.** The initial spectrum from GRB 080916C, averaged over 71-s time interval (red line); the estimated flux from GRB 080916C (red squares); the total background flux (black squares); the observed flux (background + GRB, blue circles) (a). The reconstructed flux from GRB 080916C (red squares) (b).

In Fig. 14, the initial and restored spectra from GRB 090902B are shown by red line and squares, respectively. In Fig. 14a, the spectra covered the peak and averaged over 2-s time interval. In Fig. 14b, the spectra covered the whole burst and averaged over 25-s time interval.

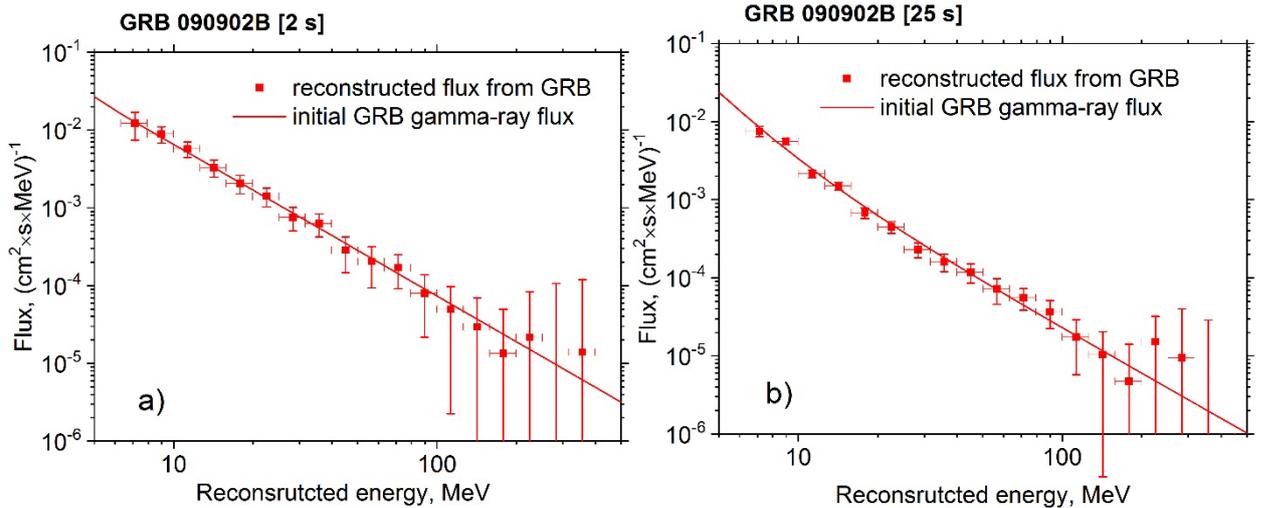

**Fig. 14.** The initial and reconstructed spectra from GRB 090902B are denoted by red line and squares, respectively. The spectra averaged over 2-s time interval (a). The spectra averaged over 25-s time interval (b).

In Fig. 15, the initial and restored spectra from GRB 090926A are shown by red line and squares, respectively. In Fig. 15a, the spectra covered one peak and averaged over 3-s time interval, covering first peak of light curve. In Fig. 15b, the spectra covered another peak and averaged over 2-s time intervals, covering second peak of light curve. In Fig. 16, the initial and restored spectra from GRB 090926A, covered the whole burst and averaged over 22-s time interval, are shown by red line and squares, respectively.

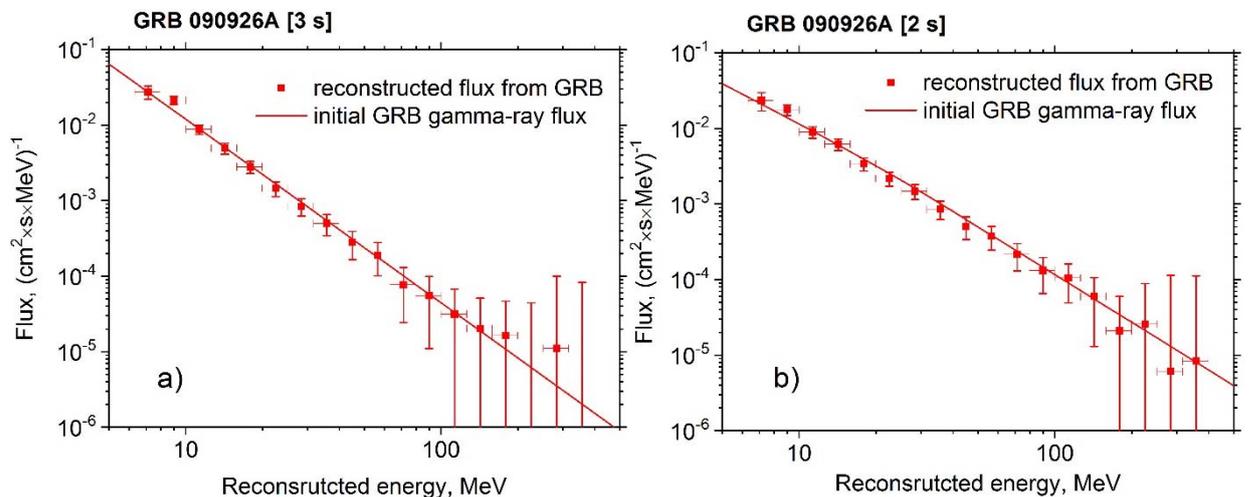

**Fig. 15.** The initial and reconstructed spectra from GRB 090926A are denoted by red line and squares, respectively. The spectra averaged over 3-s time interval, covering first peak of light curve (a). The spectra averaged over 2-s time interval, covering second peak of light curve (b).

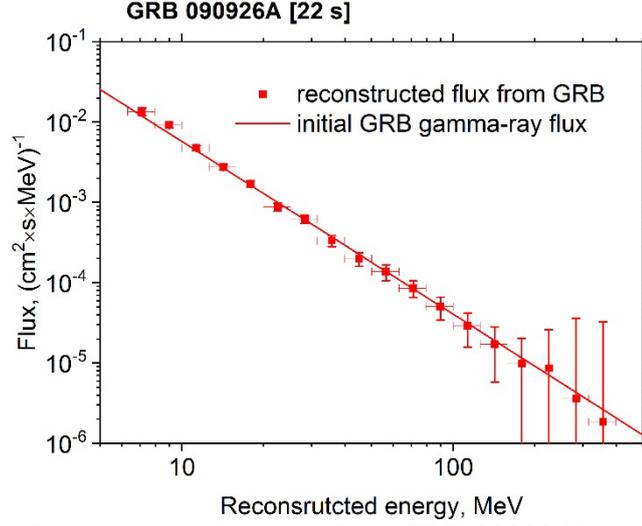

**Fig. 16.** The initial and reconstructed spectra from GRB 090926A, averaged over 22-s time interval, are denoted by red line and squares, respectively.

In Fig. 17, the initial and restored spectra from short burst GRB 090510A are shown by red line and squares, respectively. In Fig. 17a, the spectra covered the peak and averaged over 1-s time interval. In Fig. 17b the spectra covered the whole burst and averaged over 8-s time interval.

We also estimated the GRB detection thresholds. For this purpose we took the average value of the spectral index of considered GRBs and calculated two amplitude coefficients $A_{MIN}$ of the minimal detectable flux $J_{MIN} = A_{MIN} \left(\frac{E_0}{E}\right)^{2.2}$ (the pivot energy $E_0 = 1\ MeV$), integrated over the energy range of 10–100 MeV, applying TS criterion. Specifically, we defined the latter through a typical log-likelihood ratio as (Rolke, Lopez, Conrad, 2005):

$$TS = 2ln\left(\frac{L_1}{L_0}\right) \quad (6),$$

where $L_{1,0} = \lambda_{1,0}^n \frac{e^{-\lambda_{1,0}}}{n!}$ is the Poisson likelihood;
$\lambda_1$ is the mean expected number of photons from GRB and background events integrated over our energy range of interest 10-100 MeV;
$\lambda_0$ is the number of background events only;
$n$ is the total observed number of events from the simulation.

Substituting $J_{MIN} = A_{MIN}\left(\frac{E_0}{E}\right)^{2.2}$ into $\lambda_1$, applying the typical detection criteria TS = 25 (a reference mark to estimate, for example, point source sensitivity in Fermi-LAT experiment, https://www.slac.stanford.edu/exp/glast/groups/canda/lat_Performance.htm) and solving (6) for $A_{MIN}$, we obtained the following integral detection thresholds. The first amplitude $(A_{MIN})_1 = 0.42\ [MeV \times cm^2 \times s]^{-1}$ was calculated for 1-s time interval as representative of short GRBs. The second amplitude $(A_{MIN})_{10} = 0.12\ [MeV \times cm^2 \times s]^{-1}$ was calculated for 10-s time interval as representative of long GRBs. As an example of a particular GRB case, we calculated the minimal detectable flux for GRB 090510A for 1-s time interval and presented it in Fig. 17a by the black line.

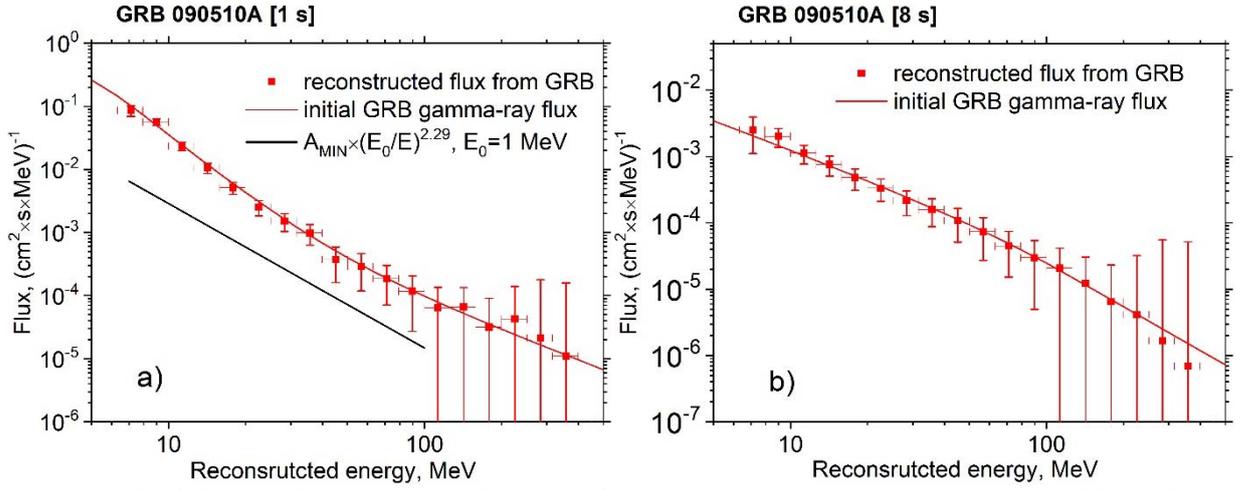

**Fig. 17.** The initial and reconstructed spectra from short GRB 090510A are denoted by red line and squares, respectively. The spectra are averaged over 1-s time interval. The minimal flux estimated for 1-s time interval with TS = 25 criterion is presented by black line (a). The spectra are averaged over 8-s time interval (b).

In above description the capability of the GAMMA-400 instrument to detect GRBs from lateral directions was analyzed for normal incidence. To consider other angles, the special simulations were fulfilled, and results of these simulations are presented in Fig. 18a. In this figure, the effective area for 50-MeV gamma rays versus angle of incidence is presented. The direction Z coincides with the direction of gamma-ray telescope axis for the main aperture. Black line corresponds to flat flux from left direction (left simulation plane, Fig. 3), red line corresponds to flat flux from top direction (top simulation plane, Fig. 3). It is necessary to mention that for both simulations the criterions (1) and (2) were applied in numerical algorithm. It is seen that influence of fluxes from the top direction is insignificant in comparison with impact of fluxes from lateral direction. For gamma rays from the top-down direction, another trigger for the main top-down aperture should be used (Leonov et al., 2019a), but this is not the point to consider in present paper. Figure 18a shows the dependence of the effective area vs angle of incidence. For the value of the effective area at hail-maximum we obtained an estimation of $\theta_{max} = 72.5°$ deg for opening angle of the cone for the lateral aperture (Fig. 18b). However, as mentioned in section 3, due to large amount of satellite platform matter located under the gamma-ray telescope, the directions from top hemisphere are allowed only. As a result, the value field of view of lateral aperture of GAMMA-400 can be estimated as $\Omega_{FoV} = 2\pi cos(\pi/2 - \theta_{max}) \approx 6$ sr.

As the structure of CC2 from CsI(Tl) columns provides only 2D projection information, and taking into account that criterion (2) is developed for each of four sides separately, the location is determined only with $\sim\pi/4$ accuracy. Therefore, the GRB optical and X-ray information, which comes from other missions, is required.

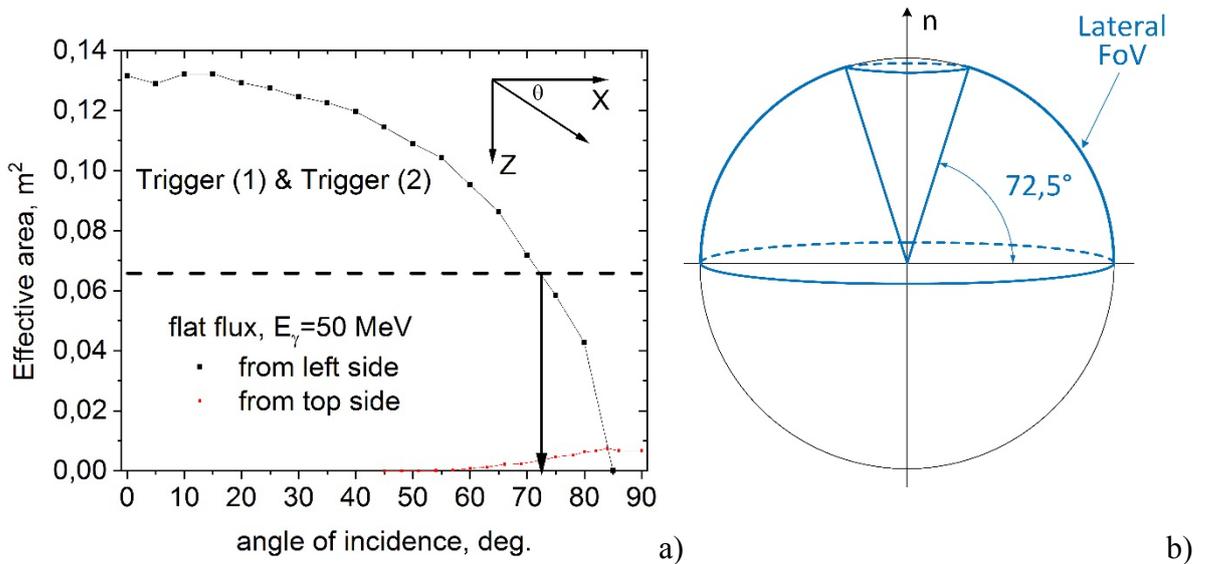

**Fig. 18.** The effective area for 50-MeV gamma rays versus angle of incidence. Black line corresponds to flat flux from left direction (left simulation plane, Fig. 3), red line corresponds to flat flux from top direction (top simulation plane, Fig. 3). For both lines the criterions (1) and (2) were applied in numerical algorithm. Value of the effective area at half-maximum gives an estimation of 72.5° for opening of a cone for lateral aperture (a). Lateral aperture FoV relative to the gamma-ray telescope axis (b).

In this section, the principle capability of the GAMMA-400 instrument to detect GRBs from lateral directions and measure their spectrum from 10 MeV to 100 MeV, when analyzing time scale of the peak or of the total period of GRBs, is reported. The GRB intensity in given time channel, while analyzing its temporal structure, will determine the possibility of investigations. To study the GRB temporal structure the additional information, concerning the count rate from each of used broad spectral channels, could be provided. The possibility to transmit the GRB temporal profiles (after data compressing) to the ground station is provided by information quota of 100 GB per day.

**6. Conclusion**

Currently, the GAMMA-400 launch is scheduled for 2030. In this paper, we showed that the capabilities of the GAMMA-400 gamma-ray telescope allow us to measure GRB spectra from the lateral aperture reliably in the energy range from ~10 to ~100 MeV. This conclusion was based on the analysis of the results of Monte-Carlo simulations. The test algorithm introduced the sample of data from space-based GBM and LAT instruments of Fermi experiment. This sample contains the brightest three long bursts: GRB 080916C, GRB 090902B, GRB 090926A and one short burst GRB 090510A. To reject background of cosmic rays the off-line additional data processing selection was constructed, providing the on-axis effective area of about 0.13 m$^2$ for each of four lateral directions. The total effective field of view for the GAMMA-400 lateral aperture is ~6 sr.

The nature of the high-energy emission and its possible connection with the low-energy classical gamma-ray emission is still debatable. There are several signs of at least bimodal behavior of high-energy emission (separate emission mechanism vs extension of low-energy emission to high energy). The detailed investigation of GRBs by the GAMMA-400 lateral aperture in energy range 10-100 MeV, being the transition zone between these emission components, could shed light on the problem.

**Acknowledgments**

The authors would like to thank the anonymous referees for their many constructive comments and suggestions that definitely helped to improve the quality of this article. The authors also thank the Editors for their management of the review.

This work was partially supported by the Russian State Space Corporation ROSCOSMOS (contract no. 024-5004/16/224), the Ministry of Science and Higher Education of the Russian Federation under Project "Fundamental problems of cosmic rays and dark matter" (contact no. 0723-2020-0040), NRNU MEPhI Academic Excellence Project (contract 02.a03.21.0005) and was performed using resources of NRNU MEPhI high-performance computing center.

**References**

Abdollahi, S., Acero, F., Ackermann, M., et al., 2020. Fermi Large Area Telescope Fourth Source Catalog. ApJS 247:33, 37pp.

Ackermann, M., Ajello, M., Atwood, W.B., et al., 2012. Fermi-LAT Observations of the Diffuse γ-Ray Emission: Implications for Cosmic Rays and the Interstellar Medium. Astrophys. J., 750:3, 1-35.

Adriani, O., Barbarino, G.G., Bazilevskaya, G.A., et al., 2011. PAMELA Measurements of Cosmic-Ray Proton and Helium Spectra. Science, 332:6025, 69-72.

Adriani, O., Barbarino, G.G., Bazilevskaya, G.A., et al., 2015. Time Dependence of the e-Flux Measured by PAMELA During the 2006 July – 2009 December Solar Minimum. Astrophys. J., 810:142, 1-13.

Ajello, M., Arimoto, M., Axelsson, M., et al., 2019. A Decade of Gamma-Ray Bursts Observed by Fermi-LAT: The Second GRB Catalog. Astrophys. J., 878:52, 1-61.

An, Q., Asfandiyarov, R., Azzarello, P., et al., 2019. Measurement of the cosmic ray proton spectrum from 40 GeV to 100 TeV with the DAMPE satellite. Sci Adv 5:eaax3793, 1-10.

Bisschoff, D., Potgieter, M.S., and Aslam, O.P.M., 2019. New Very Local Interstellar Spectra for Electrons, Positrons, Protons, and Light Cosmic Ray Nuclei. Astrophys. J., 878:59, 1-8.

Brun, R., Rademakers, F., 1997. ROOT: An Object Oriented Data Analysis Framework. Nucl. Inst. & Meth. A, 389, 81-86.

Cash, W., 1979. Parameter Estimation in Astronomy Through Aplication of the Likelihood Ratio., *Astrophys. J.*, 228, 939-947.

Galper, A.M., Topchiev, N.P., and Yurkin, Yu.T., 2018. GAMMA-400 Project. Astronomy Reports, 62 (12), 882-889.

Gruber, D., Goldstein, A., Weller von Achlefeld, V., et al., 2014. The Fermi GBM Gamma-Ray Burst Spectral Catalog: Four Years of Data. Astrophys. J. Suppl. Ser. 211:12, 1-27.

Joshi, J.C., and Razzaque, S., 2021. Modelling synchrotron and synchrotron self-Compton emission of gamma-ray burst afterglows from radio to very-high energies. Monthly Notices of the Royal Astronomical Society, *stab1329*.

Martucci, M, Munini, R., Boezio, M., et al., 2018. Proton Fluxes Measured by the PAMELA Experiment from the Minimum to the Maximum Solar Activity for Solar Cycle 24. Astrophys. J. Lett., 854:L2, 1-8.

Mikhailova, A.V., Bakaldin, A.V., Chernysheva, I.V., et al., 2020. Capabilities of the GAMMA-400 gamma-ray telescope for lateral aperture. IOP Conf. Ser.: J. Phys.: Conf. Series 1690, 012026.

Ngobeni, M.D., Aslam, O.P.M., Bisschoff, D., et al., 2020. The 3D numerical modeling of the solar modulation of galactic protons and helium nuclei related to observations by PAMELA between 2006 and 2009. Astrophys. Space Sci., 365:182, 1-18.

Leonov, A.A., Galper, A.M., Bonvicini, V., et al., 2015. Separation of electrons and protons in the GAMMA-400 gamma-ray telescope. Advances in Space Research, 56, 1538-1545.

Leonov, A.A., Galper, A.M., Topchiev, N.P., et al., 2019. Capabilities of the Gamma-400 Gamma-ray Telescope for Observation of Electrons and Positrons in the TeV Energy Range. Physics of Atomic Nuclei, 82, 6, 855-858.

Leonov, A. A., Galper, A.M., Topchiev, N.P., et al., 2019a. Multiple Coulomb scattering method to reconstruct low-energy gamma–ray direction in the GAMMA-400 space-based gamma–ray telescope. Advances in Space Research, 63, 3420-3427.

Picozza, P., Galper, A.M., Castellini, G., et al., 2007. PAMELA – A payload for antimatter matter exploration and light-nuclei astrophysics. Astroparticle Physics, 27:4, 296–315.

Poolakkil, S., Preece, R., Fletcher, C., et al., 2021. The Fermi-GBM Gamma-Ray Burst spectral catalog: 10 yr of data. Astrophys. J., 913:60, 1-20.

Rolke, W.A., Lopez, A.M., Conrad, J., 2005. Limits and confidence intervals in the presence of nuisance parameters, Nucl. Instrum. Methods Phys. Res. A, 551, 493-503

Suchkov. S.I., Galper, A.M., Arkhangelskiy, A.I., et al., 2021. Calibrating the prototype calorimeter for the GAMMA-400 γ-ray telescope on the positron beam at the Pakhra accelerator. Instruments and Experimental Techniques, 64, 5, 669-675.

Topchiev, N.P., Galper, A.M., Bonvicini, V., et al., 2016. The GAMMA-400 gamma-ray telescope for precision gamma-ray emission investigations. IOP Conf. Ser.: J. Phys.: Conf. Series 675, 032009.

Topchiev, N.P., Galper, A.M., Bonvicini, V., et al., 2016a. Perspectives of the GAMMA-400 space observatory for high-energy gamma rays and cosmic rays measurements. IOP Conf. Ser.: J. Phys.: Conf. Series 675, 032010.

Topchiev, N.P., Galper, A.M., Arkhangelskaja, I.V., et al., 2019. Space-based GAMMA-400 mission for direct gamma- and cosmic- ray observations. IOP Conf. Ser.: J. Phys.: Conf. Series 1181, 012041.

Topchiev, N.P., Galper, A.M., Arkhangelskaja, I.V., et al., 2019a. High-energy gamma- and cosmic-ray observations with future space-based GAMMA-400 gamma-ray telescope. EPJ Web of Conferences 208, 14004.

https://fermi.gsfc.nasa.gov/ssc/data/access/lat/BackgroundModels.html (the data were accessed on January, 2021).

http://fermi.gsfc.nasa.gov/ssc/data/analysis/rmfit/ (the data were accessed on January, 2021).

https://www.slac.stanford.edu/exp/glast/groups/canda/lat_Performance.htm (the data were accessed on May, 2021).

http://www.ssl.berkeley.edu/ipn3/masterli.html (the data were accessed on May, 2021).